\documentclass[10pt,conference,a4paper]{IEEEtran}
\usepackage{graphicx}
\usepackage{booktabs}
\usepackage{colortbl}
\usepackage{xcolor}
\usepackage{tikz}
\usepackage{amsmath}
\usepackage{cite}
\usepackage{enumitem}
\usepackage{placeins}
\usepackage{float}

\definecolor{AVHAccent}{RGB}{47,79,103}

\definecolor{AVHChallenge}{RGB}{0,0,0}

\definecolor{AVHResult}{RGB}{235,242,247}

\definecolor{AVHAnswer}{RGB}{48,91,125}
\definecolor{AVHAnswerBg}{RGB}{241,247,251}

\colorlet{AVHTeal}{AVHAccent}

\newcommand{\avhchallenge}[1]{%
  \par\smallskip
  \noindent
  \textbf{\textcolor{AVHChallenge}{#1}}\hspace{0.3em}%
}

\newcommand{\avhstage}[2]{%
  \par\smallskip
  \noindent
  \textbf{\textcolor{AVHAccent}{Stage~#1:} #2}
  \par
}

\newcommand{\avhnum}[1]{%
  \tikz[baseline=(char.base)]{
    \node[
      circle,
      draw=AVHAccent,
      inner sep=1pt,
      font=\tiny\bfseries
    ] (char) {#1};
  }%
}

\newcommand{\avhanswer}[2]{%
  \par\smallskip
  \noindent
  \begingroup
  \setlength{\fboxrule}{0.6pt}%
  \setlength{\fboxsep}{5pt}%
  \fcolorbox{AVHAnswer}{AVHAnswerBg}{%
    \parbox{\dimexpr\linewidth-2\fboxsep-2\fboxrule\relax}{%
      \textbf{\textcolor{AVHAnswer}{Answer to #1.}}%
      \hspace{0.3em}#2%
    }%
  }%
  \endgroup
  \par\smallskip
}

\setlist[itemize]{
  leftmargin=1.4em,
  itemsep=1pt,
  topsep=2pt
}

\title{AgenticVulHunter: Bringing Local Agentic Vulnerability Hunting to Your Terminal}

\author{
\IEEEauthorblockN{
Nishant Abhishek\IEEEauthorrefmark{1},
Ping Charoenwet\IEEEauthorrefmark{2},
Kla Tantithamthavorn\IEEEauthorrefmark{1}
}

\IEEEauthorblockA{
\IEEEauthorrefmark{1}Monash University, Australia.
\IEEEauthorrefmark{2}University of Melbourne, Australia
}

}
\usepackage[hidelinks]{hyperref}

\begin{document}

\maketitle

\begin{abstract}

Finding vulnerabilities early, before they can affect the production system, is important for reducing the risk of any software attacks. However, detecting vulnerabilities during secure code review is still difficult because the vulnerable code logic may depend on lines found across different repository files. Recent approaches such as AgenticSCR have shown that LLM agents can explore repository files and use security knowledge to improve secure code review.

However, these approaches mainly rely on external commercial LLMs and can provide a large amount of security knowledge to the agent at once. This creates an important question of whether agentic secure code review can also be performed effectively using locally deployed open-weight LLMs with more focused and relevant security knowledge.
In this paper, we propose AgenticVulHunter, an agentic secure code review approach and tool that can run with a locally deployed open-weight LLM. 

We evaluate AgenticVulHunter on 144 vulnerability cases from SCRBench using Qwen3-Coder-30B and DeepSeek V3.1 Terminus. We measure what percentage of generated review comments correctly identify the vulnerable location, match the vulnerability type, and provide a relevant security review comment together. AgenticVulHunter achieves an accuracy of 18.4\% with Qwen3-Coder-30B and 23.8\% with DeepSeek V3.1 Terminus, compared with 11.8\% for our AgenticSCR-style reproduction using Qwen3-Coder-30B and 17.5\% for AgenticSCR using Claude Sonnet~4.

Tool demo: (https://www.youtube.com/watch?v=UZw4NJ7YSoA)

\end{abstract}

\section{Introduction}

Software vulnerabilities are weaknesses in software systems that can be exploited by attackers and can affect the confidentiality, integrity, and availability of a system.  Existing secure code review tools such as CodeQL, Semgrep, and Snyk can analyse source code and report possible vulnerabilities. More recently, AgenticSCR \cite{charoenwet2026agenticscr} showed that LLM agents can explore repository files and use CWE and SAST knowledge to reason about vulnerabilities. Although this improves secure code review capabilities, several challenges still remain.

\avhchallenge{First, providing a large amount of security knowledge can introduce noise, making it harder to focus on the most relevant vulnerability knowledge.} 
AgenticSCR is designed to augment domain-specific security knowledge by incorporating a large amount of Common Weakness Enumeration (CWE) and Static Application Security Testing (SAST) knowledge into the agentic memory. However, not all of this knowledge is useful or relevant to every pull request. Recently, Yang et al. \cite{yang2025context} found that useful program context can improve vulnerability detection,  while excessive context can make it harder for the model to focus on relevant information.

\avhchallenge{Second, commercial LLMs may not be suitable for enterprise environments where source code cannot be shared externally.} AgenticSCR is designed for commercial large language models (LLMs), such as Claude, which may limit the applicability of such tools in the enterprise context. In enterprise environments, source code is confidential and cannot be sent to an externally hosted commercial model. Thus, little is known about the performance of agentic secure code review on the open-weight language models.

\avhchallenge{Third, there is a lack of practical agentic secure code review tools that developers can use before creating a pull request.} AgenticSCR shows the potential of using agents for secure code review, but it is mainly presented as a research approach rather than a tool that developers can install and use. Existing tools such as CodeQL, Semgrep, and Snyk Code can detect possible security issues~\cite{charoenwet2024sast}. However, these tools mainly rely on static analysis rather than allowing an agent to explore relevant repository files.

\section{AgenticVulHunter}

To address the above challenges, AgenticVulHunter is designed around three main ideas.  First, instead of giving all CWE and SAST knowledge to the model at once, AgenticVulHunter uses security knowledge search. It uses keywords from the collected repository knowledge to search for the closest matching CWE, and changes the search keywords for a better search result. Second, AgenticVulHunter supports locally deployed open-weight LLMs, so when a local model is used, the source code does not need to be sent to an external LLM provider.  Third, similar to previous research on terminal-based agentic code review~\cite{charoenwet2026agentic}, AgenticVulHunter provides a practical tool that allows developers to detect potential vulnerabilities directly from the terminal before creating a pull request.

\subsection{Running the Review.}
\begin{figure}[!htbp]
    \centering
    \includegraphics[width=\columnwidth]{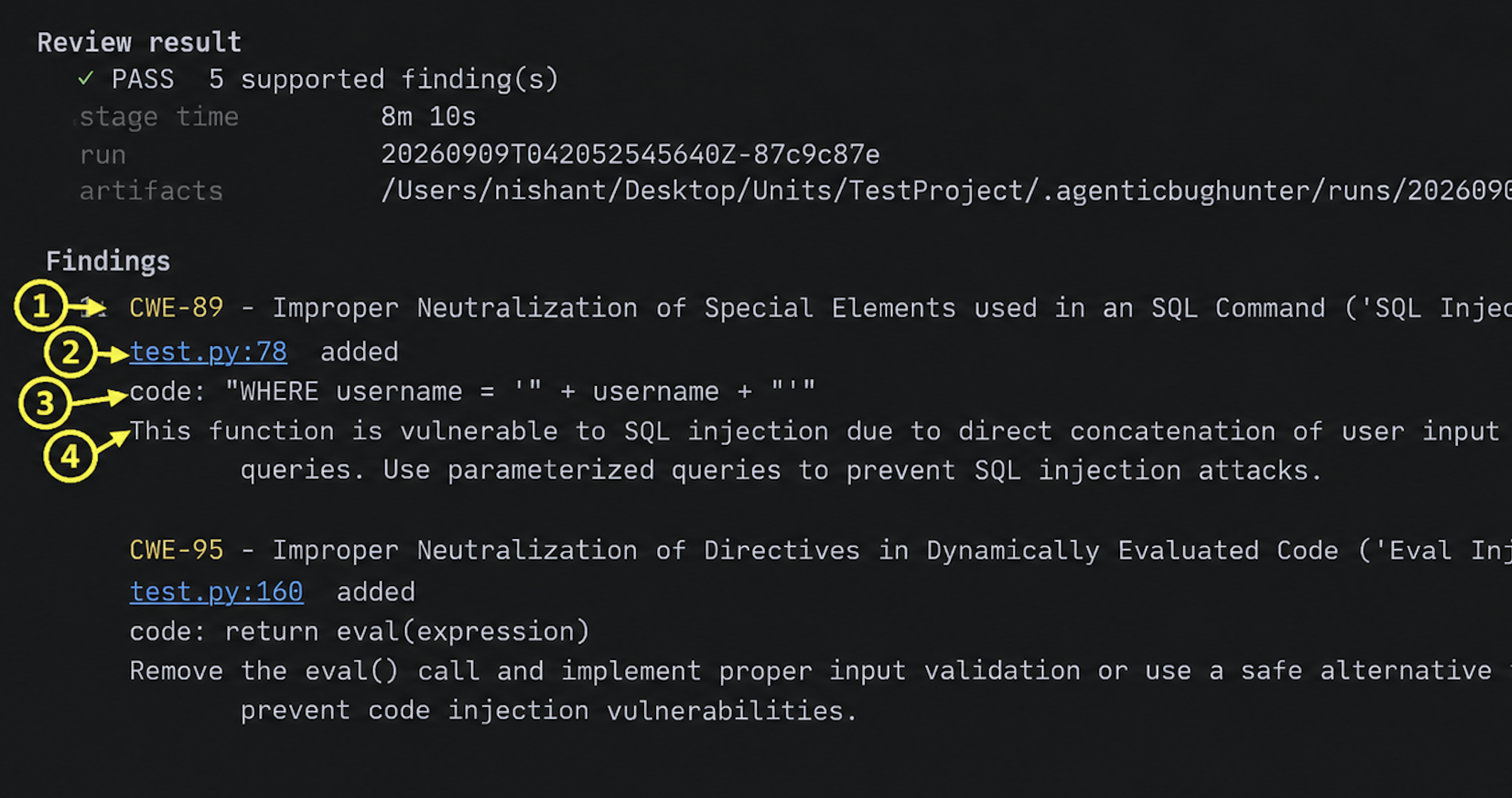}
    \caption{AgenticVulHunter review result in the command-line interface.}
    \label{fig:agenticvulhunter-cli}
\end{figure}

The developer first installs the tool using
(\textcolor{AVHTeal}{\texttt{\$ pipx install agenticvulhunter}}),
then initializes AgenticVulHunter for their working project using
(\textcolor{AVHTeal}{\texttt{\$ agenticvulhunter init}}),
and starts the secure code review using
(\textcolor{AVHTeal}{\texttt{\$ agenticvulhunter review}}).
The review checks the code diff between the latest commit and the previous commit and analyses the lines that were added or changed. Figure~\ref{fig:agenticvulhunter-cli} shows the final result.

\noindent
\avhnum{1}~\textbf{CWE Type:} the identified CWE type for the vulnerability.\\
\avhnum{2}~\textbf{Localized Line:} the file and line number where the vulnerability is found.\\
\avhnum{3}~\textbf{Vulnerable Code:} the code snippet related to the vulnerability.\\
\avhnum{4}~\textbf{Review:} the generated secure code review comment explaining the issue and how it can be fixed.

\subsection{AgenticVulHunter Architecture Design}

\begin{figure}[!htbp]
    \centering
    \includegraphics[width=0.95\columnwidth]{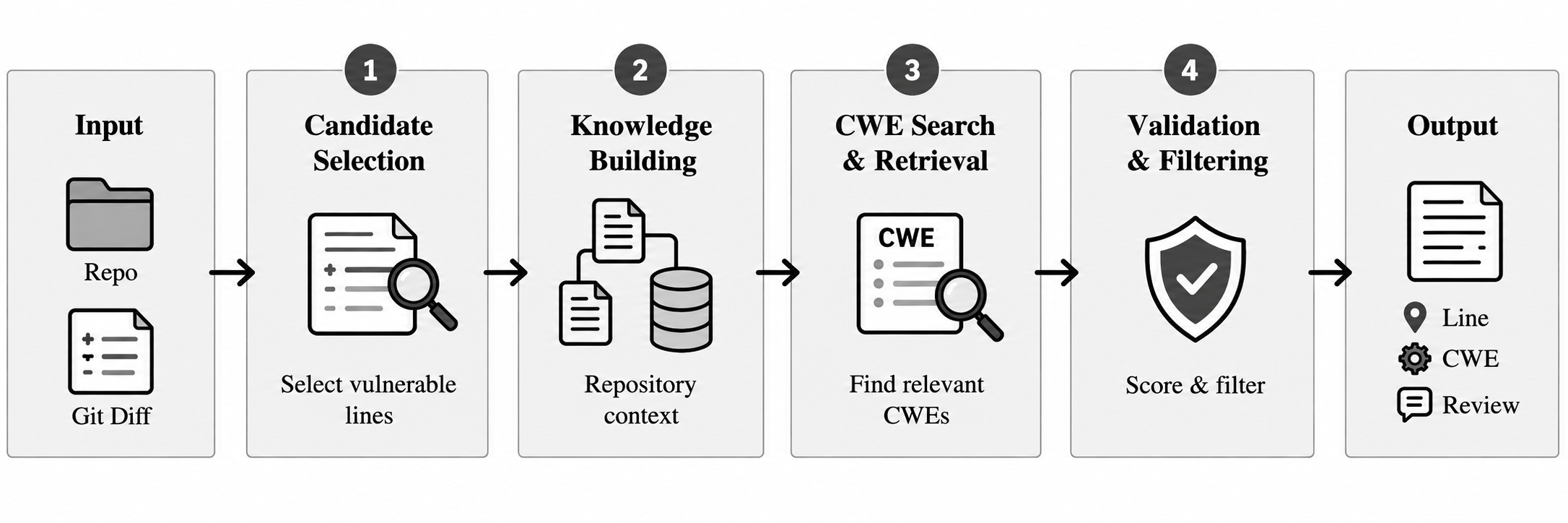}
    \caption{Overview of the AgenticVulHunter workflow.}
    \label{fig:AgenticVulHunter-architecture}
\end{figure}

AgenticVulHunter analyses the changed code through four main stages as shown in Figure~\ref{fig:AgenticVulHunter-architecture}.

\avhstage{1}{Candidate Selection}
Stage~1 (Fig.~\ref{fig:AgenticVulHunter-architecture}) analyses the Git diff, which contains the lines of code added or changed between the latest commit and the previous commit. The agent identifies changed lines that may be security-sensitive or potentially vulnerable and selects them as vulnerability candidates. The agent also has access to the repository and can explore relevant files to gather additional context when needed.

\begin{itemize}
\item \textbf{Selection Reason:} Why the candidate was selected as a potentially vulnerable line. 
\item \textbf{Confidence Score:} The model's confidence that the selected line may be vulnerable.
\item \textbf{Context Lines:} The 15 lines before and after the selected line, which are used as the starting context for Stage~2.
\end{itemize}

This information also helps the validator understand why the candidate was selected.

\avhstage{2}{Knowledge Building}
The behaviour of a potentially vulnerable candidate line may depend on code found in other parts of the repository. As shown in Figure~\ref{fig:AgenticVulHunter-architecture}, Stage~2 uses the information collected in Stage~1 to build security knowledge for each candidate line. It collects:

\begin{itemize}
\item \textbf{Function Name:} The function or method where the vulnerability candidate line is located.

\item \textbf{Operation Type:} The type of operation being performed, e.g. SQL query.

\item \textbf{Variable Source:} Where the important variables linked to the possible vulnerability come from.

\item \textbf{Vulnerability Reach:} The security-sensitive operation that the candidate may reach.

\item \textbf{Security Check:} Whether any important security check or validation is missing for the candidate line.

\item \textbf{Overall Summary:} A summary of the candidate and how it behaves.
\end{itemize}

This collected information forms the \textbf{security knowledge base} for each candidate and is passed to the later stages.

\avhstage{3}{CWE Search \& Retrieval}
Stage~3 (Fig.~\ref{fig:AgenticVulHunter-architecture}), uses the security knowledge base to search for possible vulnerability types. Based on this knowledge, the agent creates a keyword query for the BM25-based CWE retriever \cite{robertson2009bm25}. The retriever searches the CWE knowledge base and returns the most relevant CWEs for the candidate line. The agent then checks whether the retrieved CWE information matches the behaviour of the candidate.

\phantomsection
\label{sec:stage4-validation}
\avhstage{4}{Validation \& Filtering}
The vulnerable line and CWE can potentially be false positives. Therefore, Stage~4 (Fig.~\ref{fig:AgenticVulHunter-architecture}) checks each candidate and CWE pair one at a time. Stage~4 receives each candidate and its associated CWE after the BM25 search. The verifier then assigns a score between 0 and 1 for the following categories:

\begin{itemize}
\item \textbf{CWE Match:} How well the CWE matches the candidate.
\item \textbf{Security Control:} whether an important protective security check is missing in the code.
\item \textbf{Connection:} How well the variable source is connected to the possible vulnerable candidate line.
\item \textbf{Diff Causality:} How strongly the changed line introduces the vulnerability.
\item \textbf{Concrete Impact:} How strongly the candidate can lead to a real security impact.
\end{itemize}

The final score is calculated as the average of the five categories:

\begin{equation}
S =
\frac{
S_{\text{CWE}} +
S_{\text{control}} +
S_{\text{connection}} +
S_{\text{causality}} +
S_{\text{impact}}
}{5}
\label{eq:stage4-score}
\end{equation}

For our experiment, we provide an equal weight for each category to mitigate any potential bias or threshold subjectivity in our experiment. 

\section{Experimental Setup}
For answering our research question, we use the SCRBench dataset for the experimental setup and evaluation \cite{charoenwet2026agenticscr}. SCRBench is designed for pre-commit secure code review and contains 144 vulnerability cases. Following the AgenticSCR setup, GPT-4.1 is used as the judge for fair comparison.


\begin{itemize}
\setlength{\itemsep}{1pt}
\setlength{\parskip}{0pt}
\setlength{\parsep}{0pt}

    \item \textbf{RQ1:} How effectively does AgenticVulHunter perform when compared to the state-of-the-art Agentic AI for secure code review (AgenticSCR)?

    \item \textbf{RQ2:} What is the impact of the validation threshold on the performance of AgenticVulHunter?
    
\end{itemize}

\subsection{Evaluation Measures}
To answer each research question, we use the following SCRBench measures~\cite{charoenwet2026agenticscr}.
\begin{itemize}
\setlength{\itemsep}{1pt}
\setlength{\parskip}{0pt}
\setlength{\parsep}{0pt}

\item \textbf{Location Match (L):}
Percentage of comments in the same file and within $\pm 5$ lines of the ground-truth vulnerable line.

\item \textbf{Comment Type Match (T):}
Percentage of comments whose predicted CWE belongs to the same family-level CWE category as the ground-truth vulnerability.

\item \textbf{Comment Relevance (R):}
Percentage of comments whose CWE explanation correctly describes the ground-truth vulnerability.

\item \textbf{Overall Correctness (L\&R\&T):}
Percentage of comments that simultaneously satisfy Location Match, Comment Relevance, and Comment Type Match.

\end{itemize}

\subsection{Baseline Comparison}

To establish a fair comparison with the prior study, we compare the performance of AgenticVulHunter with our AgenticSCR-style reproduction using Qwen3-Coder-30B, the published AgenticSCR result using Claude Sonnet 4, Static LLM, CodeQL, Semgrep, and Snyk.

\begin{table}[!ht]
  \centering
  \caption{AgenticVulHunter and baseline results on SCRBench.}
  \label{tab:baselines}
  \scriptsize
  \setlength{\tabcolsep}{2.5pt}
  \renewcommand{\arraystretch}{1.05}

  \resizebox{\columnwidth}{!}{
  \begin{tabular}{@{}llccccc@{}}
  \toprule
  \textbf{Approach} &
  \textbf{Model} &
  \textbf{\#Cmt} &
  \textbf{L\&R\&T} &
  \textbf{L} &
  \textbf{T} &
  \textbf{R} \\
  \midrule

  \rowcolor{AVHResult}
  \textbf{AgenticVulHunter ($t=0.6$)}
  & \textbf{Qwen3-Coder-30B}
  & \textbf{201}
  & \textbf{18.4\%}
  & \textbf{38.3\%}
  & \textbf{51.2\%}
  & \textbf{62.7\%} \\

  \rowcolor{AVHResult}
  \textbf{AgenticVulHunter ($t=0.6$)}
  & \textbf{DeepSeek V3.1 Terminus}
  & \textbf{181}
  & \textbf{23.8\%}
  & \textbf{42.5\%}
  & \textbf{59.7\%}
  & \textbf{62.4\%} \\

  AgenticSCR-style
  & Qwen3-Coder-30B
  & 127
  & 11.8\%
  & 22.0\%
  & 49.6\%
  & 35.4\% \\

  \midrule

  AgenticSCR
  & Claude Sonnet 4
  & 211
  & 17.5\%
  & 30.8\%
  & 70.1\%
  & 63.5\% \\

  Static LLM
  & Claude 4.5
  & 652
  & 6.9\%
  & 28.4\%
  & 44.0\%
  & 46.5\% \\

  CodeQL
  & --
  & 1,092
  & 0.6\%
  & 3.3\%
  & 19.2\%
  & 7.2\% \\

  Semgrep
  & --
  & 376
  & 3.7\%
  & 5.1\%
  & 74.7\%
  & 26.3\% \\

  Snyk
  & --
  & 318
  & 0.9\%
  & 1.6\%
  & 39.0\%
  & 13.2\% \\

  \bottomrule
  \end{tabular}
  }

  \vspace{1mm}
  \footnotesize
  \#Cmt denotes generated review comments.
\end{table}

\section{Experimental Results}

In this section, we present the results of AgenticVulHunter for the proposed research questions.

\subsection*{\textbf{
RQ1: How effectively does AgenticVulHunter perform when compared to the state-of-the-art Agentic AI for secure code review (AgenticSCR)?
}}

\textbf{Approach.}
To analyse RQ1, we evaluate AgenticVulHunter from three perspectives. First, we compare AgenticVulHunter with our AgenticSCR-style reproduction using the same Qwen3-Coder-30B model to compare performance under the same language model. Second, we compare AgenticVulHunter with static analysis baselines, including CodeQL, Semgrep, and Snyk. Third, we evaluate AgenticVulHunter with DeepSeek V3.1 Terminus to examine whether the approach remains effective with another open-weight language model.

\textbf{Finding 1: AgenticVulHunter with targeted security knowledge retrieval outperforms the AgenticSCR-style reproduction using the same model.}
The current state-of-the-art approach, AgenticSCR, populates the LLM memory with a large amount of CWE and SAST knowledge. This contains approximately 38,615 lines of security knowledge, including 28,888 lines of CWE knowledge covering 944 CWE entries and 9,727 lines of SAST rules. In contrast, AgenticVulHunter starts from the changed line and builds security knowledge about the vulnerability candidate through agentically exploring relevant repository files. This collected knowledge is then used to search for CWE information relevant to the current vulnerability candidate line. This is designed to provide more relevant information for reasoning without overwhelming smaller models with excessive security knowledge. As shown in Table~\ref{tab:baselines}, AgenticVulHunter achieves an L\&R\&T score of 18.4\% with Qwen3-Coder-30B, compared with 11.8\% for the AgenticSCR-style reproduction using the same model, representing an improvement of 6.6 percentage points.

\textbf{Finding 2: AgenticVulHunter outperforms the static analysis baselines.} Current static analysis tools such as CodeQL, Semgrep, and Snyk rely mainly on predefined rules and analysis patterns to identify possible vulnerabilities. In contrast, AgenticVulHunter allows the agent to explore relevant files and collect additional context before generating the review. Table~\ref{tab:baselines} shows that AgenticVulHunter achieves an L\&R\&T score of 18.4\%, compared with 0.6\% for CodeQL, 3.7\% for Semgrep, and 0.9\% for Snyk.

\textbf{Finding 3: AgenticVulHunter is effective with another open-weight language model.}
To examine whether the approach remains effective beyond Qwen3-Coder-30B, we evaluate AgenticVulHunter using DeepSeek V3.1 Terminus. As shown in Table~\ref{tab:baselines}, AgenticVulHunter achieves an L\&R\&T score of 23.8\% with DeepSeek V3.1 Terminus, compared with 18.4\% using Qwen3-Coder-30B. The DeepSeek result is also higher than the 17.5\% reported by AgenticSCR using Claude Sonnet~4. This shows that the overall AgenticVulHunter approach remains effective when used with a different open-weight language model and is not dependent on Qwen3-Coder-30B alone.

Therefore, these findings together answer RQ1 by showing that AgenticVulHunter performs better than the AgenticSCR-style reproduction when using the same Qwen3-Coder-30B model, substantially outperforms the static analysis baselines, and remains effective when applied with another open-weight language model.

\avhanswer{RQ1}{
AgenticVulHunter improves the L\&R\&T score from 11.8\% to 18.4\% using Qwen3-Coder-30B, an increase of 6.6 percentage points compared with our AgenticSCR-style reproduction using the same open-weight language model.}

\subsection*{\textbf{RQ2: What is the impact of the validation threshold on the performance of AgenticVulHunter?}}

\textbf{Approach.}
To analyse RQ2, we vary the Stage~4 validation threshold from $t=0.0$ to $t=1.0$. We analyse how changing the validation threshold affects the L\&R\&T score.

\begin{figure}[H]
    \centering
    \includegraphics[width=\columnwidth]{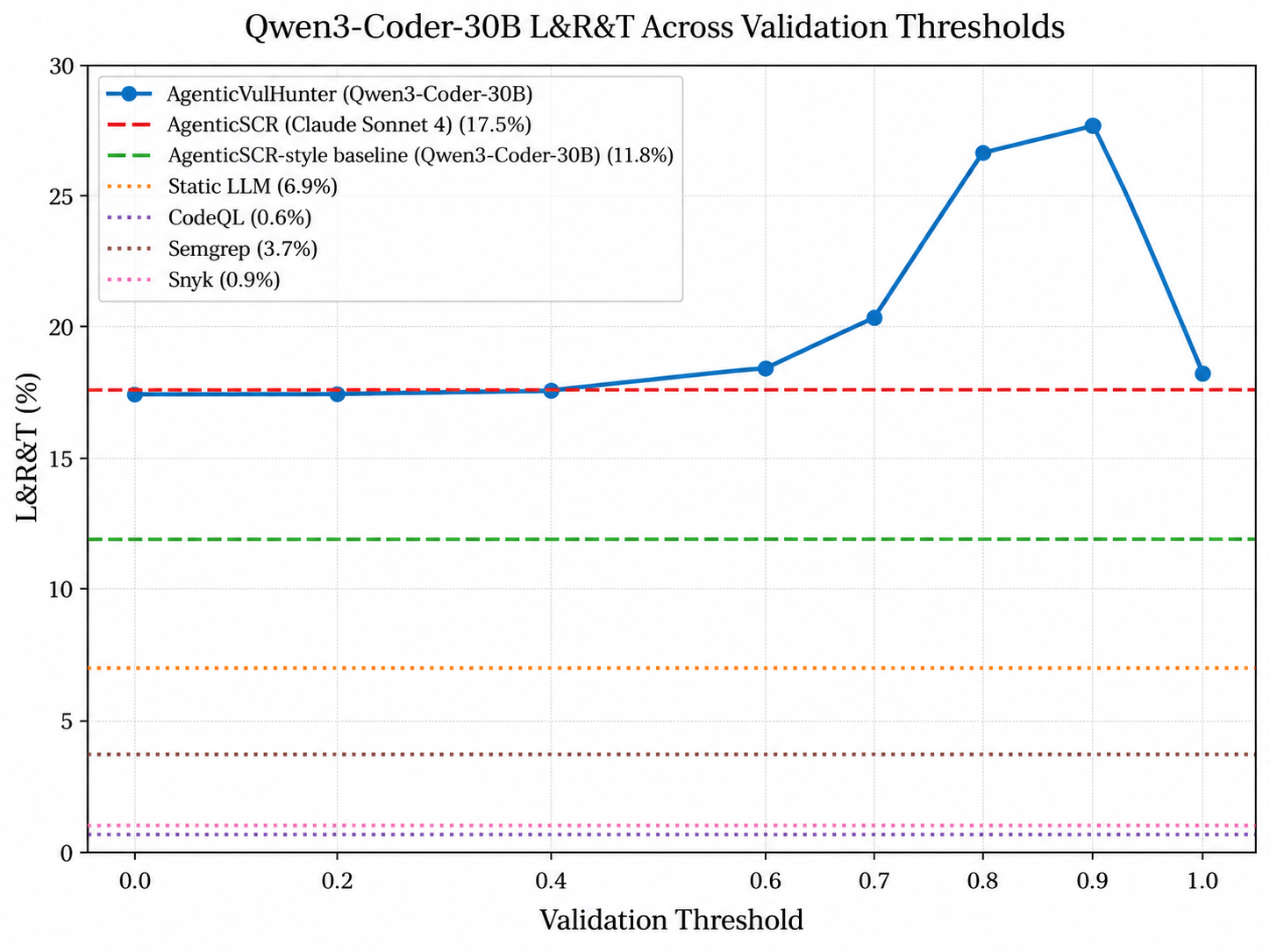}
    \caption{L\&R\&T performance of AgenticVulHunter across validation thresholds.}
    \label{fig:rq2}
\end{figure}

\textbf{Finding 4: Increasing the validation threshold generally improves the L\&R\&T score by filtering weaker vulnerability line candidate reviews.}
As shown in Fig.~\ref{fig:rq2}, For AgenticVulHunter using Qwen3-Coder-30B, the L\&R\&T score increases from 17.4\% at $t=0.0$ to its highest value of 27.6\% at $t=0.9$. This behaviour is supported by the scoring mechanism used in AgenticVulHunter. In Stage~4, the model uses the evidence collected from previous stages to score each vulnerability line candidate between 0 and 1 across five validation categories (exact CWE fit, security protection failure, vulnerability connection from the source of the variable to its reach, diff causality caused by the changed line, and security impact). At lower thresholds, reviews with weaker or incomplete evidence can still pass Stage~4. As the threshold increases, a review needs stronger and more consistent evidence across all validation categories to remain. As a result, higher thresholds retain a larger proportion of reviews that are correct in location, vulnerability type, and relevance together.

\textbf{Finding 5: AgenticVulHunter remains competitive across all tested validation thresholds.}
As shown in Fig.~\ref{fig:rq2}, Across all tested thresholds from $t=0.0$ to $t=1.0$, AgenticVulHunter remains above the 11.8\% AgenticSCR Qwen3-Coder-30B baseline. From $t=0.4$ onward, the L\&R\&T score is also competitive or higher than the 17.5\% reported by AgenticSCR using Claude Sonnet~4. 

Therefore, these findings together answer RQ2 by showing that increasing the validation threshold generally improves the L\&R\&T performance of AgenticVulHunter by filtering weaker reviews, while the performance remains higher than the 11.8\% AgenticSCR Qwen3-Coder-30B baseline across all tested thresholds.

\avhanswer{RQ2}{
Across validation thresholds from 0.0 to 1.0, the L\&R\&T score of AgenticVulHunter ranges from 17.4\% to 27.6\%. AgenticVulHunter remains higher than the AgenticSCR Qwen3-Coder-30B baseline across all tested thresholds, and from $t=0.4$ onward performs at least on par with or higher than the 17.5\% reported by AgenticSCR using Claude Sonnet~4.
}

\section{Threats to Validity}


\textbf{Construct Validity.}
For a fair comparison, We use GPT-4.1 as an LLM-as-a-Judge to evaluate the correctness of generated review comments, following AgenticSCR. However, LLM judges may produce different results across runs. To reduce this threat, we use the same evaluation settings as SCRBench and AgenticSCR, including a temperature of 0.

\textbf{External Validity.}
We evaluate AgenticVulHunter on 144 vulnerability cases from SCRBench, covering different open-source projects and vulnerability types. However, the results may not generalise to other programming languages, repositories, or codebases. Further evaluation on additional datasets is needed.

\section{Conclusion}
AgenticVulHunter is developed to investigate whether pre-commit agentic secure code review can work effectively with locally deployed open-weight language models. Our results show that this approach improves secure code review performance with open-weight language models. Using Qwen3-Coder-30B, AgenticVulHunter achieves an L\&R\&T score of 18.4\%, compared with 11.8\% for the AgenticSCR using the same model. These results show that locally deployable open-weight language models can support effective agentic secure code review when they are provided with relevant and focused security knowledge.  To support the reproducibility of our work, the tool and the replication package is publicly-available at \url{https://github.com/awsm-research/AgenticVulHunter/}.

\bibliographystyle{IEEEtran}
\bibliography{references}

\end{document}